\documentclass{sig-alternate}
\usepackage{amsmath}
\usepackage{amssymb}
\usepackage{mathtools}
\usepackage{algorithm}
\usepackage{algpseudocode}
\usepackage{pbox}
\usepackage{lipsum}
\usepackage{graphicx}
\usepackage{multicol}
\usepackage{caption}
\usepackage{etoolbox}
\usepackage{float}
\usepackage{graphicx}
\usepackage{subfig}

\makeatletter
\patchcmd{\maketitle}{\@copyrightspace}{}{}{}
\makeatother

\makeatletter
\newcommand*{\algrule}[1][\algorithmicindent]{\makebox[#1][l]{\hspace*{.5em}\vrule height .75\baselineskip depth .25\baselineskip}}%

\newcount\ALG@printindent@tempcnta
\def\ALG@printindent{%
    \ifnum \theALG@nested>0
        \ifx\ALG@text\ALG@x@notext
            \addvspace{-3pt}
        \else
            \unskip
            \ALG@printindent@tempcnta=1
            \loop
                \algrule[\csname ALG@ind@\the\ALG@printindent@tempcnta\endcsname]%
                \advance \ALG@printindent@tempcnta 1
            \ifnum \ALG@printindent@tempcnta<\numexpr\theALG@nested+1\relax
            \repeat
        \fi
    \fi
    }%
\usepackage{etoolbox}
\patchcmd{\ALG@doentity}{\noindent\hskip\ALG@tlm}{\ALG@printindent}{}{\errmessage{failed to patch}}
\makeatother

\DeclareMathOperator*{\argmax}{arg\,max}
\title{Improving Active Learning in Systematic Reviews}
\numberofauthors{3}
\author{
\alignauthor
Gaurav Singh\\
      \affaddr{University College London}\\
      \affaddr{London, UK }\\
      \email{gaurav.singh.15@ucl.ac.uk}
\alignauthor
James Thomas\\
      \affaddr{University College London}\\
      \affaddr{London, UK}\\
      \email{james.thomas@ucl.ac.uk}
\alignauthor
John Shawe-Taylor\\
      \affaddr{University College London}\\
      \affaddr{London, UK}\\
      \email{j.shawe-taylor@ucl.ac.uk}
}

\begin{document}
\maketitle

\begin{abstract}
 Systematic reviews are essential to summarizing the results of different clinical and social science studies. The first step in a systematic review task is to identify all the studies relevant to the review. The task of identifying relevant studies for a given systematic review is usually performed manually, and as a result, involves substantial amounts of expensive human resource. Lately, there have been some attempts to reduce this manual effort using active learning. In this work, we build upon some such existing techniques, and validate by experimenting on a larger and comprehensive dataset than has been attempted until now. Our experiments provide insights on the use of different feature extraction models for different disciplines.
 More importantly, we identify that a naive active learning  based screening process is biased in favour of selecting similar documents. We aimed to improve the performance of the screening process using a novel active learning algorithm with success. Additionally, we propose a mechanism to choose the best feature extraction method for a given review.
\end{abstract}

\section{Introduction}
Systematic reviews are essential in various domains to summarize evidence from multiple sources. They are used in the formulation of public policies and also contribute in medicine in the development of clinical guidance \cite{gough2012introduction, chalmers2002brief}. They involve the searching, screening and synthesis of research evidence from multiple sources available in the form of textual documents. It is critical in systematic reviews to identify all studies relevant to the review in order to minimise bias. 
It is easy to see that such a process can be extremely time-consuming and demand significant human resources \cite{o2015using}.

Researchers have exploited active learning text classification to make the process of systematic review more efficient. Active learning is an iterative process that starts from a small set of labelled studies and gradually learns to differentiate between relevant and irrelevant studies \cite{miwa2014reducing, wallace2010semi, cohen2006reducing}.  



Feature extraction models are essential to the success of active learning. 
\cite{wallace2010semi} proposed a multi-view approach that encoded documents in terms of different feature vectors such as, words that appear in the title and the abstract, keywords and MeSH terms. Each feature space is used to train a different classifier, and an ensemble of those classifiers predicts the final outcome based on majority voting. 

Most previous approaches worked well for studies in the clinical domain, but it was shown in \cite{miwa2014reducing} that the task of identifying relevant studies in public health domain resulted in poorer performance. It was argued that the task of identifying relevant studies is more challenging in this domain compared to clinical domain due to the presence of documents from wide ranging disciplines (e.g., social science, occupational health, education, etc.)  in public health compared to more specific studies included in clinical systematic reviews \cite{beahler2000information}. The authors proposed using topic modelling to extract topic based features using Latent Dirichlet Allocation \cite{blei2003latent}. In addition to LDA, there are a variety of different topic models available in machine learning like probabilistic latent semantic indexing (pLSI). 


Hashimoto  \emph{et al.} \cite{hashimoto2016topic} presented a topic detection model to improve the performance of the active learning classifier. It uses a neural network model, popularly referred to as paragraph vectors \cite{le2014distributed} to extract fixed length feature representations from the documents. It is argued that the paragraph vectors can encode information contained in the sequence of words. As opposed to topic modelling techniques like LDA that treat documents as bag-of-words \cite{mikolov2013distributed}. It was shown in \cite{le2014distributed} that paragraph vectors can compute semantic relatedness between different textual contents of varying lengths. LDA on the other hand treats documents as bag-of-words and does not utilise the information stored in the sequence and context of words. \cite{hashimoto2016topic} models topics using clustered representation of paragraph vectors and represents each document in terms of the normalized distance from cluster centroids.

In this work, we perform experiments on a comprehensive dataset of reviews derived from wide ranging and diverse domains. We  perform experiments with different feature extraction techniques including bag-of-words, paragraph vectors and topic modelling using LDA. We observe that different feature extraction methods work well for different domains, but no isolated feature extraction technique works well for all reviews. In previous works \cite{wallace2010semi, hashimoto2016topic}, a naive active learning algorithm is used. 
 A naive active learning algorithm suffers from the problem of retrieving studies that are similar to the previously retrieved studies. It happens because the active learning classifier does not receive enough diverse samples to classify correctly. In \cite{sharma2015active}, authors propose active learning with rationals, where, rationals are basically groups of words (phrases) that describe the label. These rationals are asked from labeler while providing the label and can lead to extra manual effort. In addition, such an approach can only work well with bag of words representation of documents, and not with paragraph vector based representation. In many cases, (including our case), we can not get additional data for the labels because of extra manual effort involved. \cite{tong2001support} proposed an active learning strategy that uses a non-linear SVM, but training such classifiers is time consuming and not suited for active learning in clinical text classification, since the system can not keep the labeler waiting for documents to labelled. 

We propose a novel active learning algorithm for clinical text classification that reduces the bias by including novelty in addition to the relevance of documents, and derive meaningful insights from our results on how to choose a feature extraction model for a particular review.
Our contributions can be summarized as follows: 
\begin{itemize}
    \item We perform experiments on the use of different feature extraction models over a comprehensive set of reviews, and contrary to the published results in \cite{hashimoto2016topic} observe on a larger dataset that paragraph vector based topic modelling does not work better compared to bag-of-words and/or simple paragraph vectors based approach for active learning in systematic reviews.
    \item We propose a novel active learning algorithm that removes the inevitable bias in naive algorithms based solely on relevance, and instead develop an algorithm that in its early phases uses the novelty of documents, 
     in addition to their relevance.
    \item We derive insights from our experiments on the performance of different feature extraction models for different domains, and propose an effective way to choose the correct feature extraction model that requires no prior knowledge of the domain and/or review.
\end{itemize}

\section{Methods}

\subsection{Feature Extraction}

\subsubsection{Bag-of-Words}
The most basic and oldest of feature extraction models from text documents is referred to as \emph{bag-of-words}. In the bag-of-words model, each of the possible terms in the entire corpus of documents is used to construct a vocabulary. 
In addition to simple bag-of-words, the terms may be weighted by tf-idf, which measures the relative popularity of different words in a given document and reduces the popularity of commonly used words in the corpus.

\subsubsection{Latent Dirichlet Analysis}
Topic modelling techniques, like LDA \cite{blei2003latent}, have been used in the area of systematic reviews to extract topical features from studies \cite{mo2015supporting}. Such topical representations extracted from the documents are then used for training the active learning classifier.

\subsubsection{Paragraph Vectors}
Recently, a neural network model has been proposed that learns word vectors and paragraph vectors (PV) in a joint manner \cite{le2014distributed}. That is in contrast to the approach of learning them separately.  Originally, word vectors represented only the words, but paragraph vectors are able to represent a sequence of words in the form of phrases, sentences or paragraphs. Paragraph vectors have been reported to work with success in some previous works \cite{hashimoto2016topic}, and are expected to encode natural language better than other topic models. 
They have been used to measure similarities between \textit{Wikipedia} article and research papers \cite{dai2015document}. They have also been used to extract topics from studies used in systematic reviews \cite{hashimoto2016topic}.

\subsubsection{Topic Modelling using PV}
In \cite{hashimoto2016topic}, a technique for topic modelling using paragraph vectors was proposed. It is argued that paragraph vectors lead to better feature extraction by jointly learning the vector representation of words and documents. Hence, clustering such paragraph vectors can lead to better topic modelling compared to topic models that are based on bag-of-words representation, like LDA \cite{blei2003latent}. 

\subsubsection{Clustering Bag-of-Words}
We can cluster the  tf-idf based bag-of-words representation of documents to obtain topics. The documents can then be represented in terms of the cluster-distance matrix. 
It is the most basic of the different topic extraction models that was used as a baseline in \cite{hashimoto2016topic}.

\subsection{Active Learning}

The active learning process begins with a small number of manually labelled documents. These documents are then used to train a classifier that can be used to differentiate relevant and irrelevant studies among the rest of the studies. These studies are ordered in decreasing probability of relevance and the top-$k$ studies are manually reviewed by an expert reviewer. These manually reviewed $k$ studies are then used to retrain the active learning classifier along with the previously labelled studies.

There is an obvious problem with such a naive active learning algorithm. The classifier can easily get biased towards studies that are chosen in the beginning of the process. Such a classifier would continue to look for similar studies based on its current knowledge. It would lead to a biased sample of studies in the training set of the classifier. We hypothesized that including novelty in addition to relevance while choosing documents for active learning can lead to an overall improvement in performance. Some other approaches like \cite{sharma2015active} require additional information about the labels, and can only work well with bag-of-words representation of documentation. In comparison, our method can work well with both bag-of-words and distributed representations (like paragraph vectors). Note, that to the best of our knowledge, naive active learning is the only algorithm used with success in systematic reviews \cite{hashimoto2016topic, o2015using}.

\subsubsection{Proposed Algorithm}
To solve the above mentioned problems, we extract topics from documents using LDA topic model. It gives as output topic vectors for each document in the corpus $v(d) \in \mathbb{R}^{k \times 1}$. These topic vectors from different documents form a matrix $V \in \mathbb{R}^{n \times k}$, where $n$ is the number of studies in the corpus and $k$ is the number of topics. We create a separate matrix of topic vectors for documents $d$ that have already been manually labelled. We refer to the  set of already labelled documents as $\mathcal{H}$, and the matrix of topic vectors for documents $d \in \mathcal{H}$ is referred to as $S \in \mathbb{R}^{|\mathcal{H}| \times k}$. We denote the set of currently unlabelled documents as $\mathcal{G}$, and set of all documents as $\mathcal{D}$. We use principal component analysis to compute the top-$t$ principal eigen vectors of $S^TS$. We define the probability of a document being novel as:

\begin{equation}
    p(n|d) = 1-\frac{\| UU^T v(d)\|_2}{\|v(d)\|_2}
\end{equation}

where $U \in \mathbb{R}^{k \times t}$ contains $t$ principal eigen vectors, and $UU^T \in \mathbb{R}^{k \times k}$ is the subspace formed by top-$t$ principal eigen vectors of $S^TS$. It basically measures the novelty of a document by projecting its topic vector on the principal subspace formed by topic vectors of documents in the training set. The projection should be small for a document to be considered novel and vice versa, if not novel. Additionally, we obtain the probability of a given document $d$ being relevant $p(r|d)$ from the classifier. We compute the probability of document being both relevant and novel as:

\begin{equation}
    p(r,n|d) =  p(r|d) *  p(n|d)
\end{equation}

We order the documents by above mentioned $p(r,n|d)$, and then select the top-$k$ documents in each iterative step of the active learning algorithm for review by an expert reviewer. We continue using novelty of a document in the active learning process till we have discovered a certain fixed number of topics. We assign a document $d$ to topic $i$ if $$i= \argmax_k~ v_k(d)$$ where $v_k(d)$ is the value in $k_{th}$ index of topic vector $v(d)$. Afterwards, we stop incorporating the novelty in the active learning process and continue solely based on relevance. We assume that a given topic has been discovered if any document in the labelled set $\mathcal{H}$ is assigned to that topic. We provide a more formal description of the above mentioned process in Algorithm \ref{ref:alg}.

\begin{algorithm}[ht!]
\caption{Systematic Reviews Learning}
\label{ref:alg}
\begin{algorithmic}[1]
\Procedure{Active Learning}{}
\State \# Get initial manually labelled set
\State $\mathcal{H}$ = \Call{getInitialLabelledSet}{}
\State $\mathcal{G} = \mathcal{D}-\mathcal{H}$ 
\State $t = 3$
\State max\_topics $ = 150$
\State $s = 25$
\State $n = 0$
\State \# Get topical representation of docs using LDA
\State $V$ = LDA(D, n\_topic=300) 
\While{$n < $max\_topics}
\State clf = Classifier()
\State clf = \Call{trainClassifier} {clf, $\mathcal{H}$}
\State $R$ = \Call{getRelevanceScores}{clf, $\mathcal{G}$} \Comment{p(r|d)}
\State $N$ =  \Call{getNoveltyScores}{$V$, $\mathcal{H}$, $\mathcal{G}$, $t$} \Comment{p(n|d)}
\For{$\forall d \in \mathcal{G}$}
\State scores(d) = $R(d)*N(d)$
\EndFor
\State \# Get top $s$ documents as per scores
\State $\mathcal{G}'=$ \Call{getTopkDocuments}{scores, s} 
\State $\mathcal{H} = \mathcal{H} \cup \mathcal{G}' $
\State $\mathcal{G} = \mathcal{G} - \mathcal{G}'$
\State \# Topics discovered in labelled set
\State $n =$ \Call{getNumberOfTopicsDiscovered}{V, $\mathcal{H}$}
\EndWhile
\While{$|\mathcal{G}| > 0$}
\State clf = Classifier()
\State clf = \Call{trainClassifier} {clf, $\mathcal{H}$}
\State $R$ = \Call{getRelevanceScores}{clf, $\mathcal{G}$} \Comment{p(r|d)}
\For{$\forall d \in \mathcal{G}$}
\State scores(d) = $R(d)$
\EndFor
\State \# Get top $s$ documents as per scores
\State $\mathcal{G}'=$ \Call{getTopkDocuments}{scores, s} 
\State $\mathcal{H} = \mathcal{H} \cup \mathcal{G}' $
\State $\mathcal{G} = \mathcal{G} - \mathcal{G}'$
\EndWhile

\EndProcedure
\Function{getNoveltyScores}{$V$, $\mathcal{H}$, $\mathcal{G}$, $t$}
\State $N=\{\}$
\State \# Get top $t$ Eigen Vectors
\State $U$ = \Call{getEigenVectors}{$\mathcal{H}$, ncomps = $t$} 
\For{$\forall d \in \mathcal{G}$}
\State $N(d) = 1 - \frac{\|UU^Tv(d) \|_2}{\|v(d)\|_2}$
\EndFor
\State \Return $N$
\EndFunction

\Function{getNumberOfTopicsDiscovered}{V, $\mathcal{H}$}
\State $T = \{\}$
\For{$\forall d \in \mathcal{H}$}
\State $t = \argmax_k v_{k}(d)$
\State $T = T \cup t$ 
\EndFor
\State \Return |T|
\EndFunction
\end{algorithmic}

\end{algorithm}

\subsection{Evaluation}
\subsubsection{Evaluation Method} 
Firstly, we evaluate the performance of active learning classifier using different feature extraction methods. We experiment with both a linear-SVM (as used in \cite{hashimoto2016topic, miwa2014reducing}) and a logistic regression classifier and observe that performance of both classifiers is very similar. Therefore, we use a logistic regression classifier that models the posterior probability of a study being eligible $p(y|d)$. 

Secondly, we evaluate our proposed active learning algorithm and compare against the naive active learning approach. The active learning process starts with a small set of manually labelled studies. Features are extracted from this labelled set using different feature extraction models mentioned above. At each step of the iterative process a fixed set of studies are reviewed manually by an expert reviewer. 
We extract a sample of top-$k$ studies at each iterative step for manual labelling from the ordered list of candidate studies \cite{cohen2006reducing, o2015using}. 

Thirdly, we derive insights from our experiments that would help us choose the correct feature extraction model without prior knowledge of the reviews.

\begin{figure*}
\subfloat[Cooking Skills]{\includegraphics[width = 6cm]{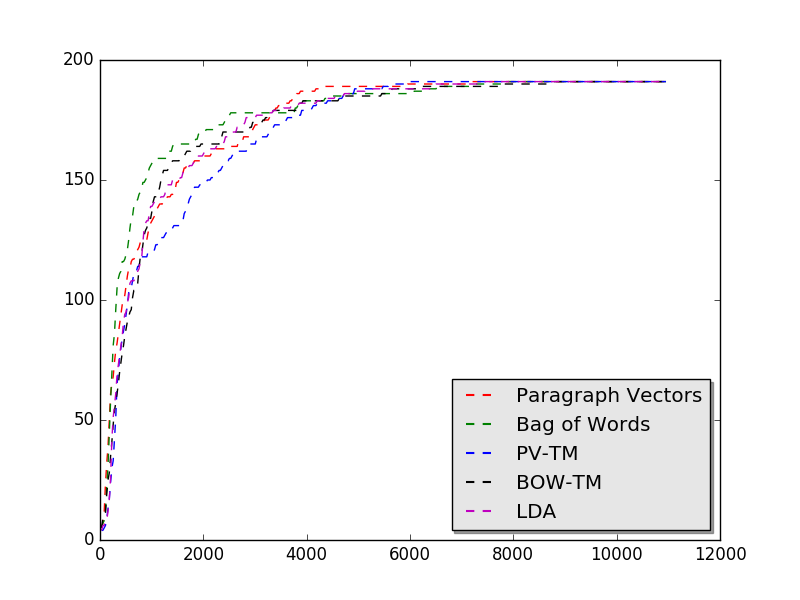}} 
\subfloat[Youth Development]{\includegraphics[width = 6cm]{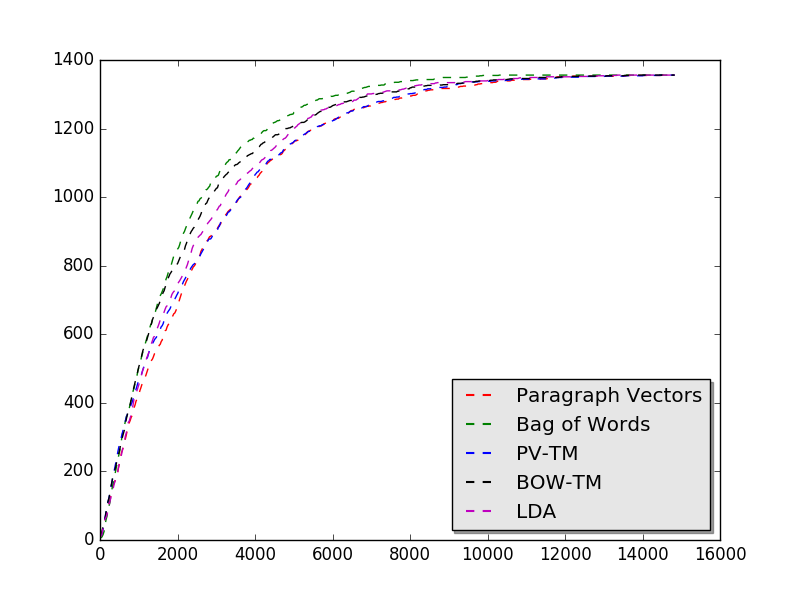}}    
\subfloat[Tobacco Packaging]{\includegraphics[width = 6cm]{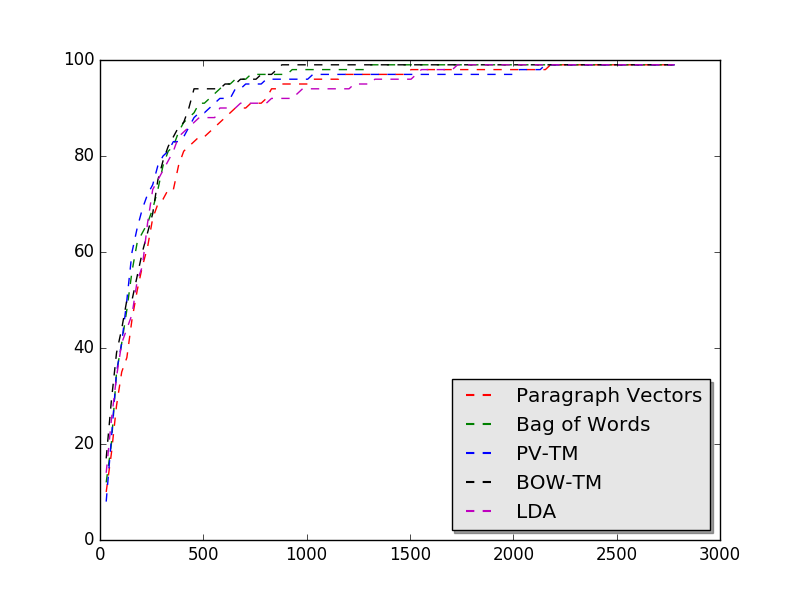}}\\
\subfloat[FABC]{\includegraphics[width = 6cm]{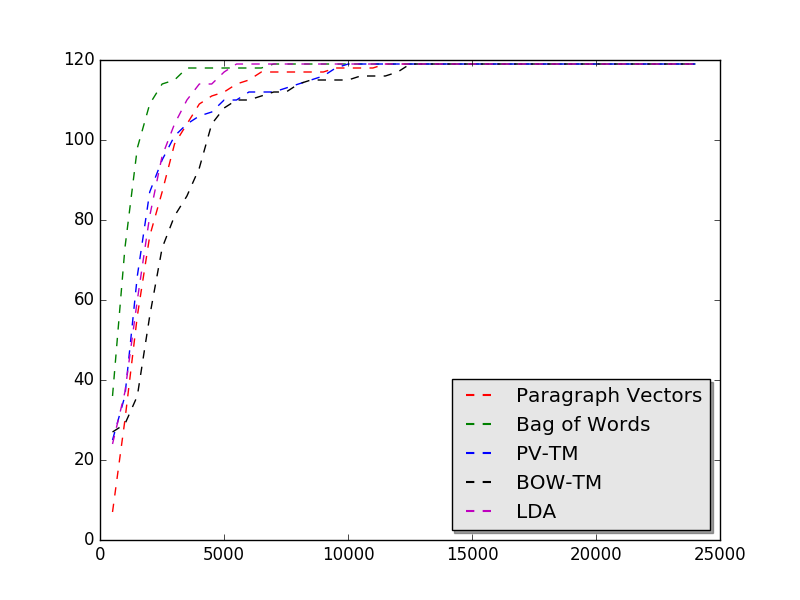}}
\subfloat[CAFO]{\includegraphics[width = 6cm]{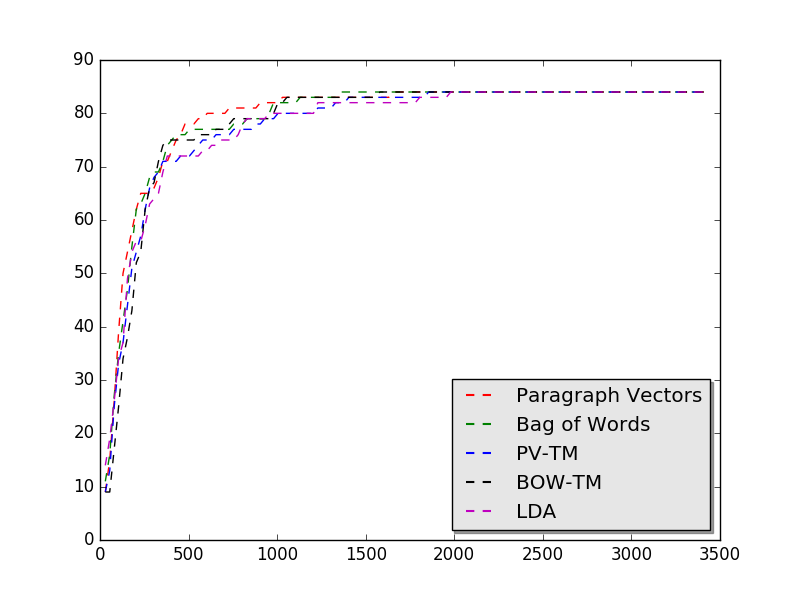}}
\subfloat[NPA]{\includegraphics[width = 6cm]{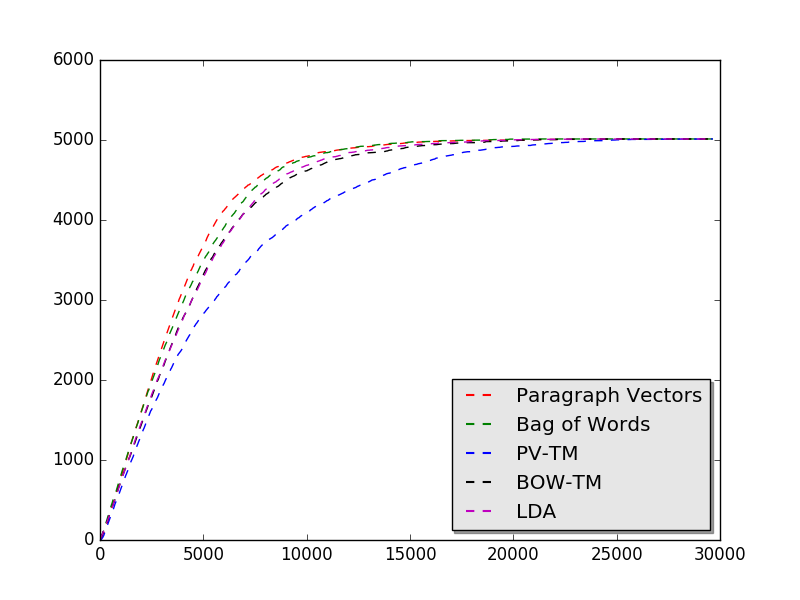}}\\
\subfloat[ASCD]{\includegraphics[width = 6cm]{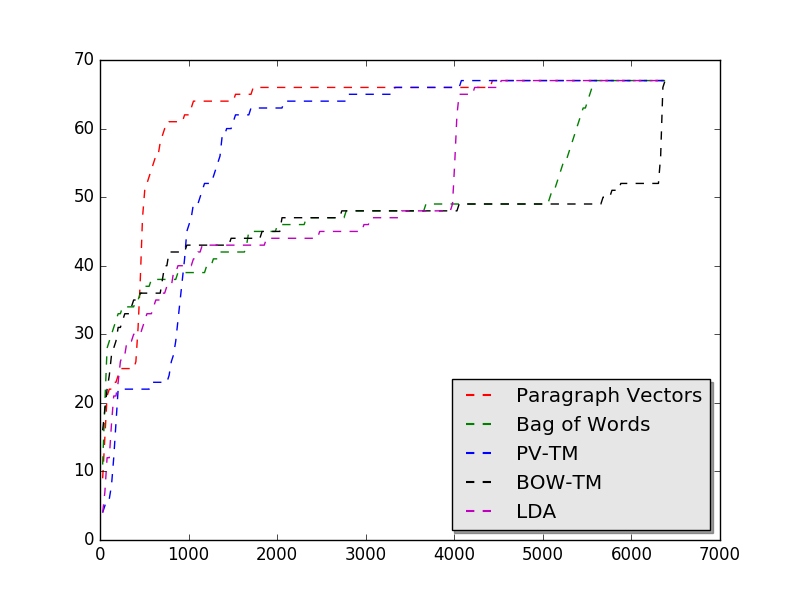}}
\subfloat[DPCAD]{\includegraphics[width = 6cm]{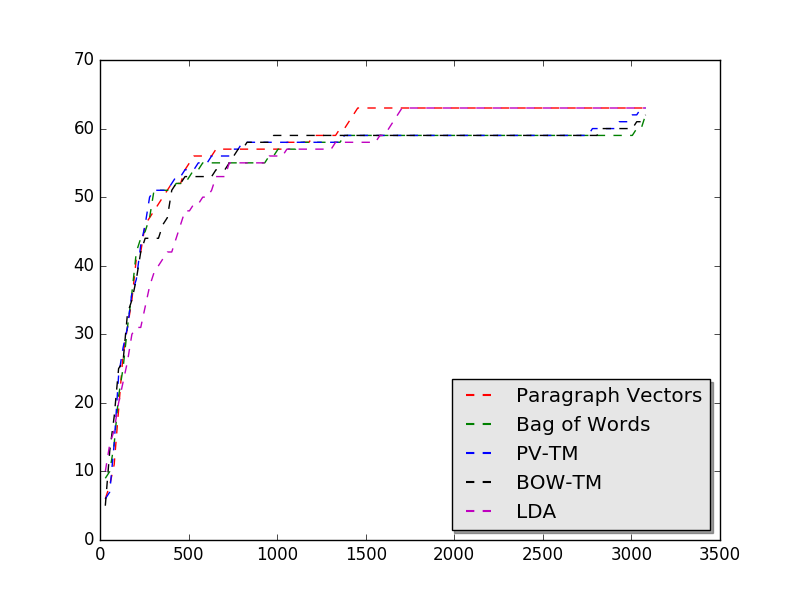}}
\subfloat[STCS]{\includegraphics[width = 6cm]{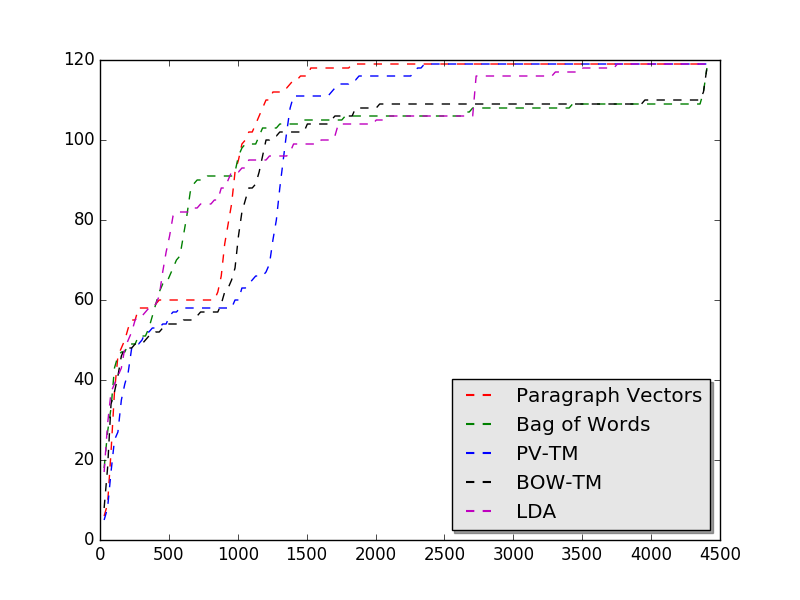}}\\
\subfloat[FVC]{\includegraphics[width = 6cm]{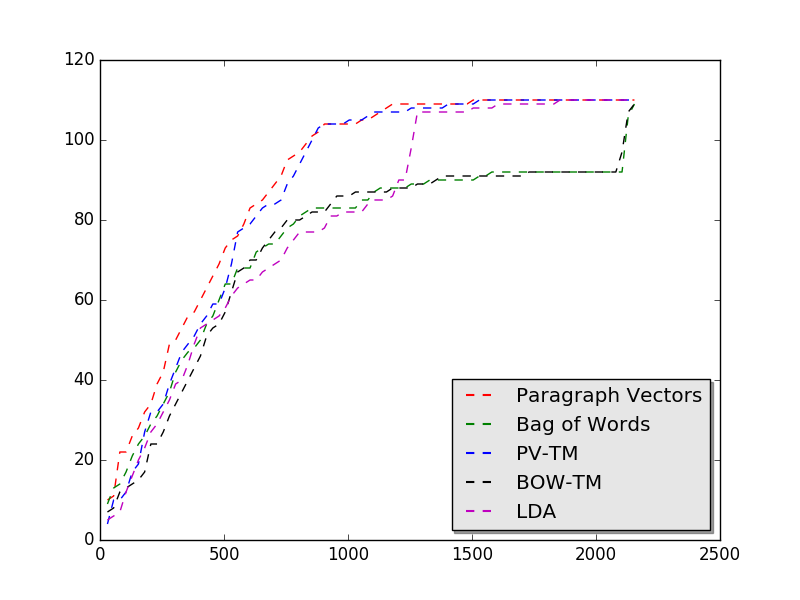}}
\subfloat[SPCHD]{\includegraphics[width = 6cm]{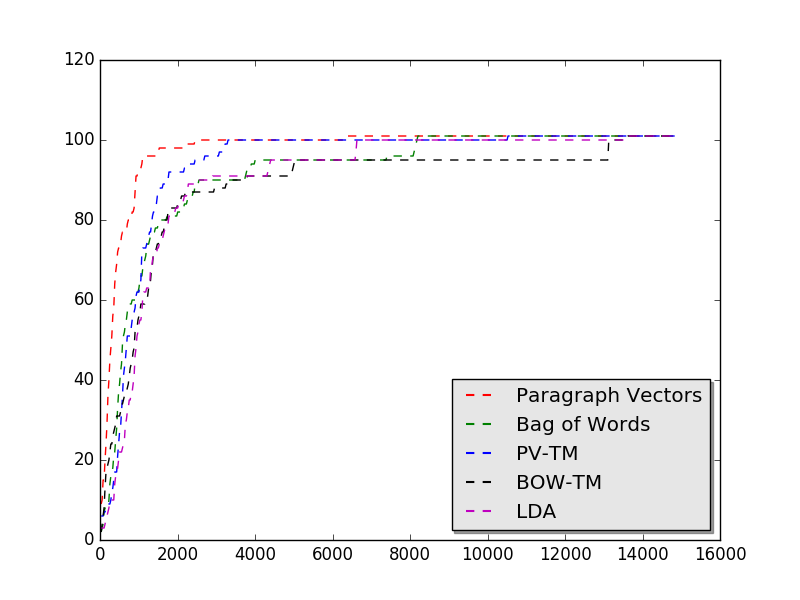}}
\subfloat[LHVS]{\includegraphics[width = 6cm]{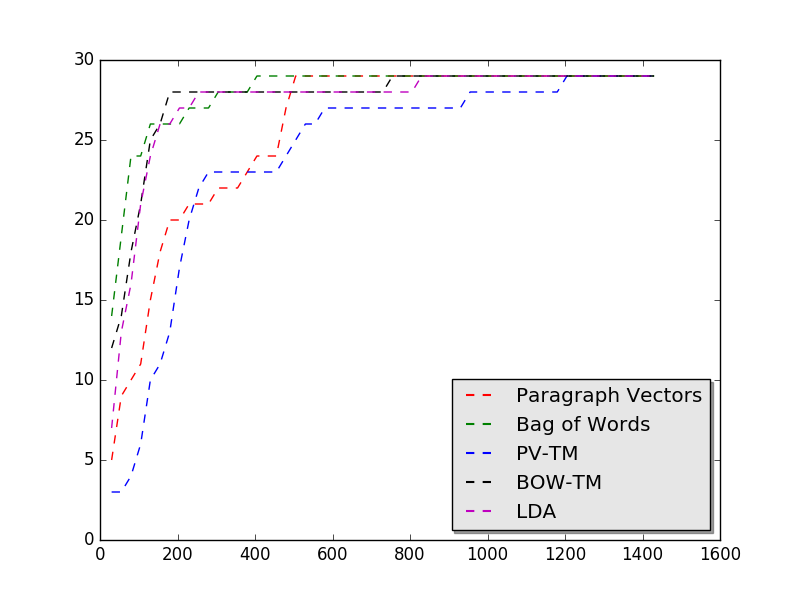}}

\caption{X-axis represents the number of documents that have been manually annotated and Y-axis represents the number of relevant documents that have been discovered. We can observe that different feature extraction methods work better for different reviews. BoW works better for public health documents and PV performs better on clinical studies. We can clearly infer that no  feature extraction method is obviously superior to others for all the documents/domains.}
\label{fig:linegraph}
\end{figure*}

 All our datasets are already labelled by expert reviewers. We simulated a human feedback active learning strategy \cite{hashimoto2016topic, miwa2014reducing, wallace2010semi}. 
Our evaluation strategy is similar to the previously published work \cite{hashimoto2016topic}.

\subsubsection{Parameter Tuning}
We tune the different parameters of PV using cross validation. We keep the number of topics in LDA at 300, as in \cite{hashimoto2016topic}. We experimented with dimensionality of paragraph vectors and found that using 300 dimensional document vectors performed better in general across different reviews. We experimented with higher value of dimension such as 1000 (as in \cite{hashimoto2016topic}), but the results were not better. We tuned the regularization parameter for the linear classifier, but observed that the results are extremely stable across different values of the regularization parameter and used the value that gave the best results. The number of principal eigen vectors that we use to compute $p(n|d)$ are 3, i.e. $t=3$ (line 42 in Algorithm \ref{ref:alg}). The value of parameter max\_topics was set to 150 in line 6 of Algorithm \ref{ref:alg}. It denotes the number of topics to be explored before we stop using novelty as a criteria in our active learning algorithm. We used the value of $s=25$ in the algorithm. A small values for $s$ implies that the classifier would have to be learned too often, and larger value  would lead to selecting too few relevant documents in the beginning, and as a result decrease in overall performance.

\subsubsection{Evaluation Metric}
We evaluate the performance of our active learning process using a metric called WSS@95. It stands for Work Saved over Sampling at 95\% yield. It can be expressed as:
\begin{equation}
    WSS@95 = (1-burden) \mid yield \geq 95\%
\end{equation}
\begin{equation}
    yield = \frac{TP^M+TP^A}{TP^M+TP^A+FN^A}
\end{equation}
\begin{equation}
    burden = \frac{TP^M+TN^M+TP^A+FP^A}{N}
\end{equation}
where N are the \# of studies and superscript M and A denote manual and automatic screening decisions. TP, FP, TN and FN stand for \# of True Positives, False Positives, True Negatives and False Negatives respectively. These notations/definitions are similar to those mentioned in \cite{hashimoto2016topic}.

\subsection{Datasets}
We used a number of public health, animal study and clinical review datasets from completed systematic reviews. Some of these datasets have been previously used in Miwa \emph{ et al.} \cite{miwa2014reducing} and Hashimoto  \emph{et al.} \cite{hashimoto2016topic}.  We summarize the characteristics of different datasets used in our experiments in Table \ref{tab:dataset}. 
\begin{table}[]
    \centering
    \resizebox{\columnwidth}{!}{%
    \begin{tabular}{c|c|c|c}
    \hline\hline
        Dataset & Domain & \pbox{20cm}{Num. of \\citations} &  \pbox{20cm}{~\\Fraction of\\relevant studies\\}\\ \hline\hline
        LHVS & Clinical & 1430 & 0.018\\
        ASCD & Clinical & 6381 & 0.010\\
        FABC & Clinical & 24469 & 0.005\\
        DPCAD & Clinical & 3087 & 0.019\\
        STCS & Clinical & 4415 & 0.026\\
        FVC & Clinical & 2157 & 0.050\\
        SPCHD & Clinical & 14841 & 0.006\\
        NPA & Animal Studies & 29659 & 0.168\\
        CAFO & Animal Studies & 3434 & 0.023\\
        Cooking Skills & Public Health & 10957 & 0.017\\
        Youth Development & Public Health & 14834 & 0.091\\
        Tobacco Packaging & Public Health &  2792 & 0.034\\ \hline
    \end{tabular}
    }
    \caption{Statistics of the systematic review datasets used in the experiment.}
    \label{tab:dataset}
\end{table}
We give a short description of different datasets used:
\begin{itemize}
\item LHVS: Leukodepletion for patients undergoing heart valve surgery
\item ASCD: Amiodarone versus other pharmacological interventions for prevention of sudden cardiac death
\item FABC: Altering availability and proximity of products for changing selection and consumption of food, alcohol and tobacco
\item NPA:  Animal studies on neuropathic pain
\item CAFO: Concentrated animal feeding operations
\item DPCAD: Psychological and pharmacological interventions for depression in patients with coronary artery disease
\item STCS: Preoperative statin therapy for patients undergoing cardiac surgery
\item FVC: Interventions for increasing fruit and vegetable consumption in children aged 5 years and under
\item SPCHD: Internet-based interventions for the secondary prevention of coronary heart disease
\item Youth Development, Cooking Skills and Plain Tobacco Packaging as described in \cite{hashimoto2016topic, miwa2014reducing} 

\end{itemize}

\section{Results}
We investigate the performance of different feature extraction techniques in terms of relevant studies discovered and amount of manual annotations required. We plot the number of relevant studies identified for a given amount of manual annotation for different reviews that range from public health to clinical science. Both these quantities converge to their maximum value during the last iteration of the active learning process. We plot the performance for different feature extraction models in Figure \ref{fig:linegraph}. We denote topic modelling based on paragraph vectors with PV-TM, BoW stands for bag-of-words, BoW-TM denotes topic modelling based on BoW model and LDA refers to Latent Dirichlet Allocation. We can see in Figure \ref{fig:linegraph} that  different feature extraction methods work well for different studies. We observe that in majority of the cases, paragraph vectors and BoW models perform better than the rest. In documents pertaining to public health, BoW performs well, whereas for clinical reviews PV performs well. These results are  contrary to previous results in \cite{hashimoto2016topic}, where the topic modelling based on PVs seemed to outperform other models across all disciplines. We observe that BoW and PV perform better than topic modelling based on paragraph vectors on most datasets. We show that it is not possible to find a single feature extraction method that performs superior to others across all domains, but we need to identify one per each review.

\begin{figure*}
\subfloat[Cooking Skills]{\includegraphics[width = 6cm]{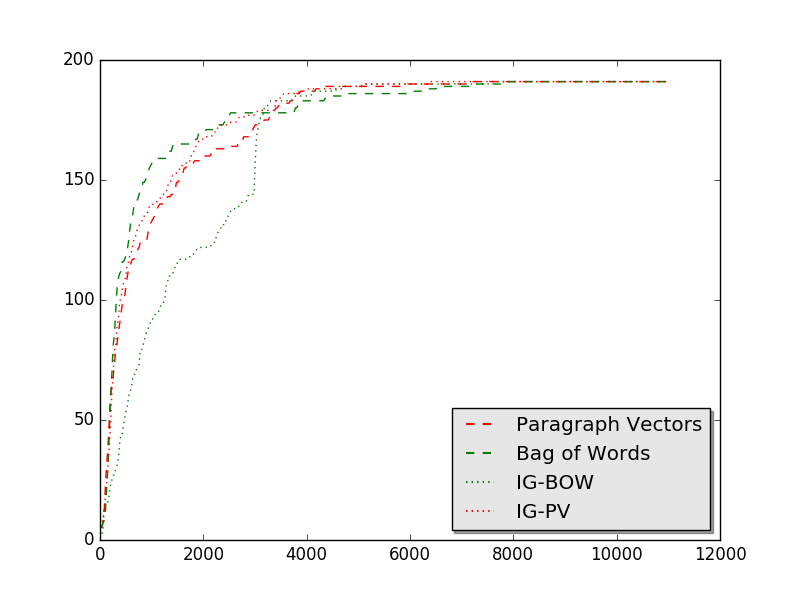}} 
\subfloat[Youth Development]{\includegraphics[width = 6cm]{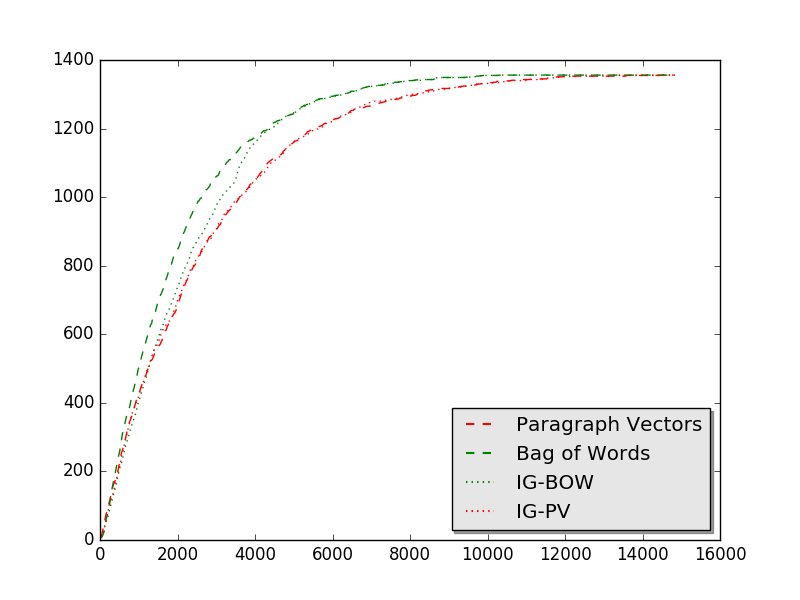}}    
\subfloat[Tobacco Packaging]{\includegraphics[width = 6cm]{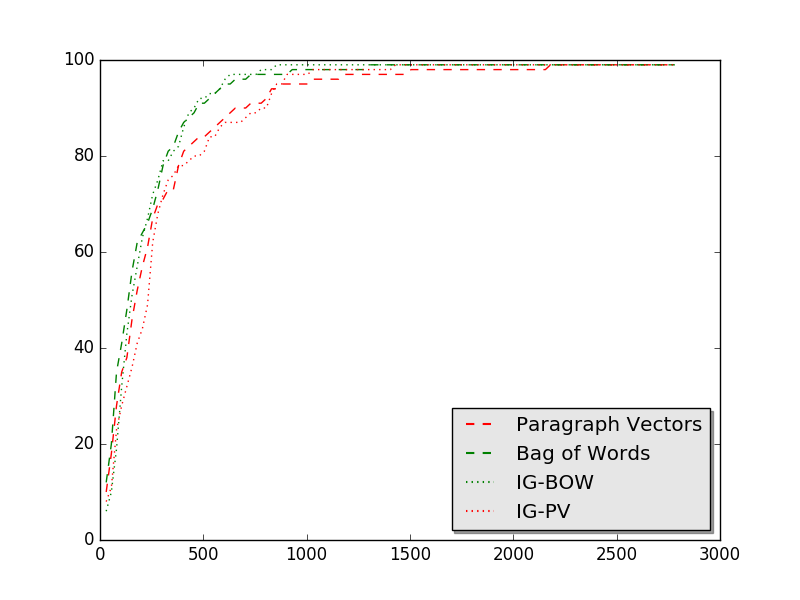}}\\
\subfloat[FABC]{\includegraphics[width = 6cm]{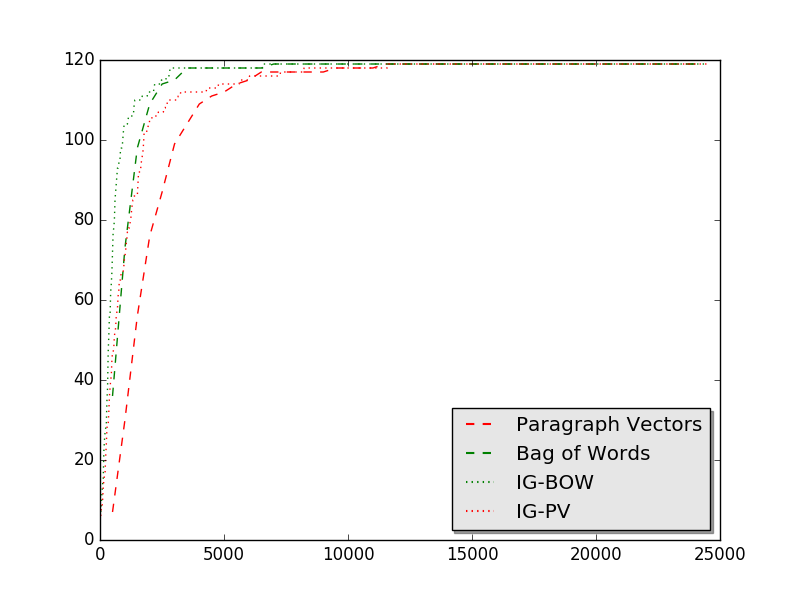}}
\subfloat[CAFO]{\includegraphics[width = 6cm]{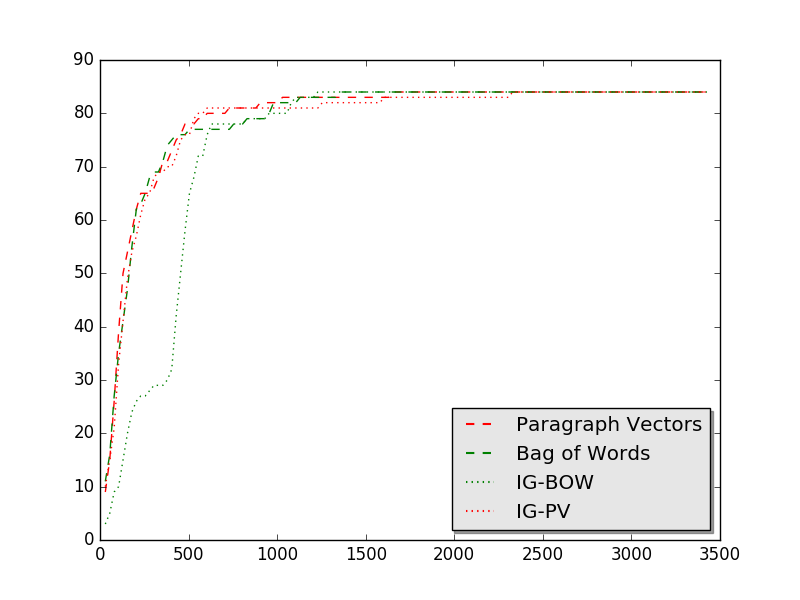}}
\subfloat[NPA]{\includegraphics[width = 6cm]{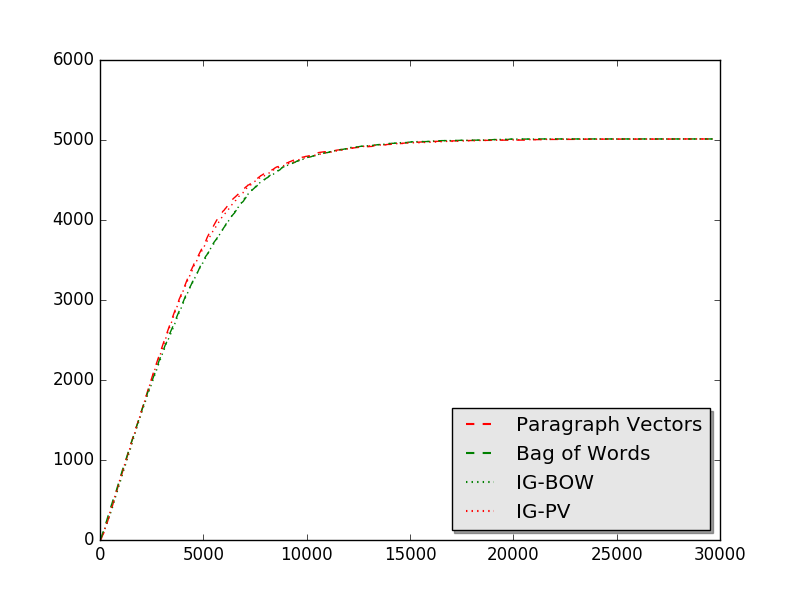}}\\
\subfloat[ASCD]{\includegraphics[width = 6cm]{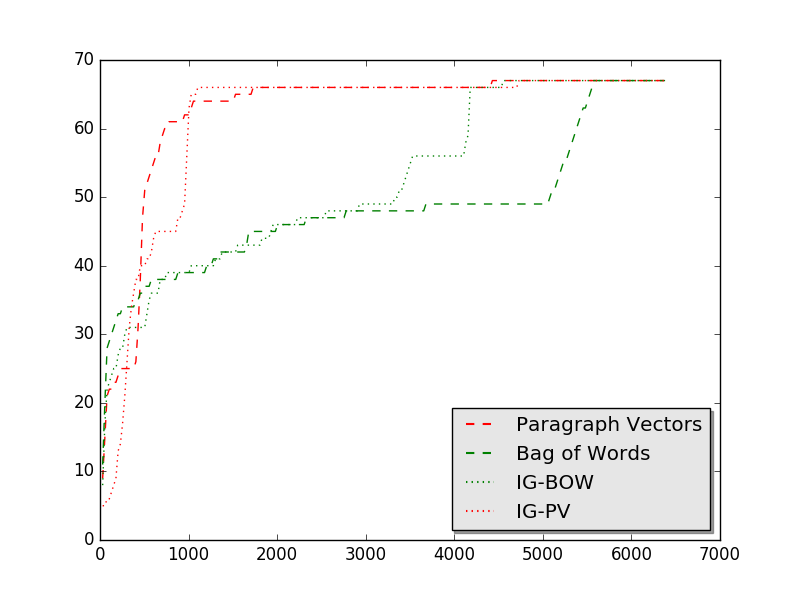}}
\subfloat[DPCAD]{\includegraphics[width = 6cm]{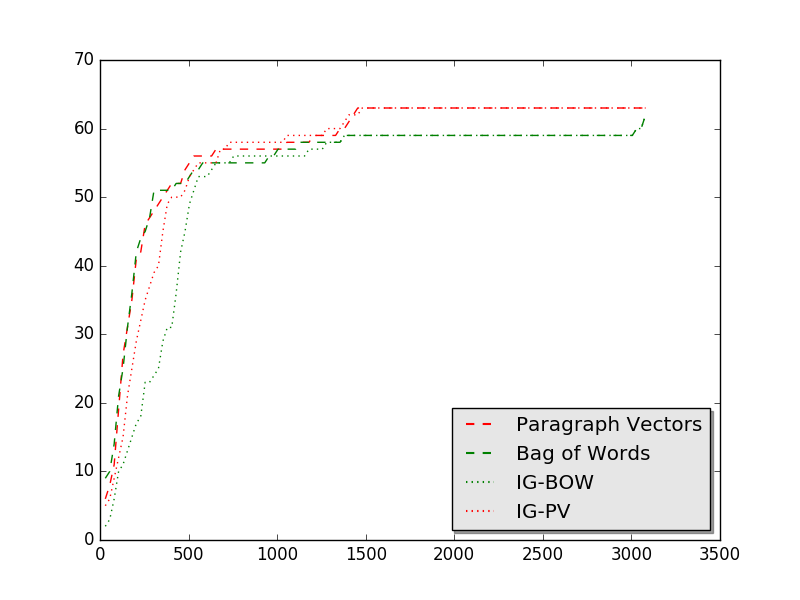}}
\subfloat[STCS]{\includegraphics[width = 6cm]{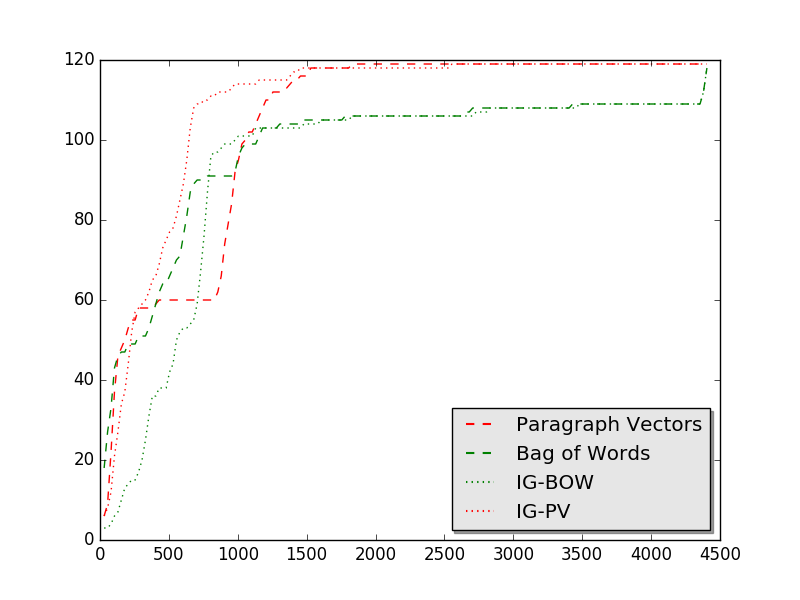}}\\
\subfloat[FVC]{\includegraphics[width = 6cm]{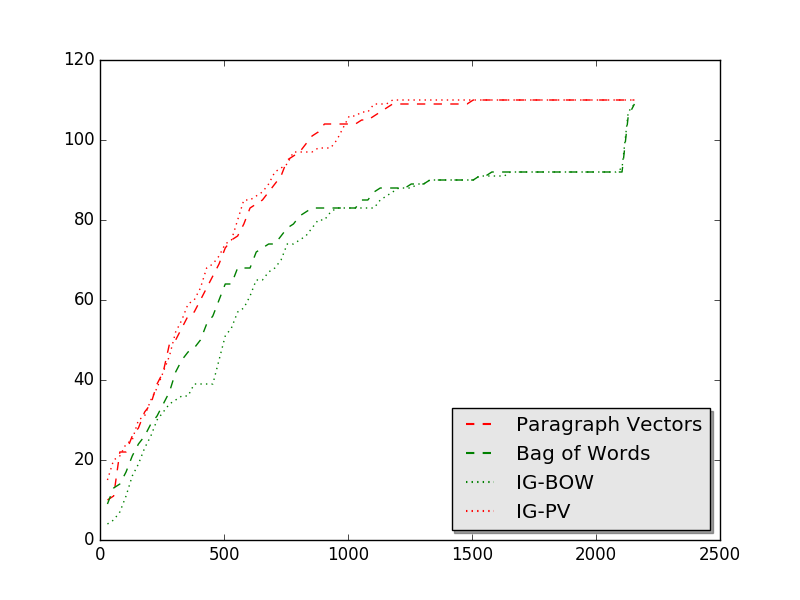}}
\subfloat[SPCHD]{\includegraphics[width = 6cm]{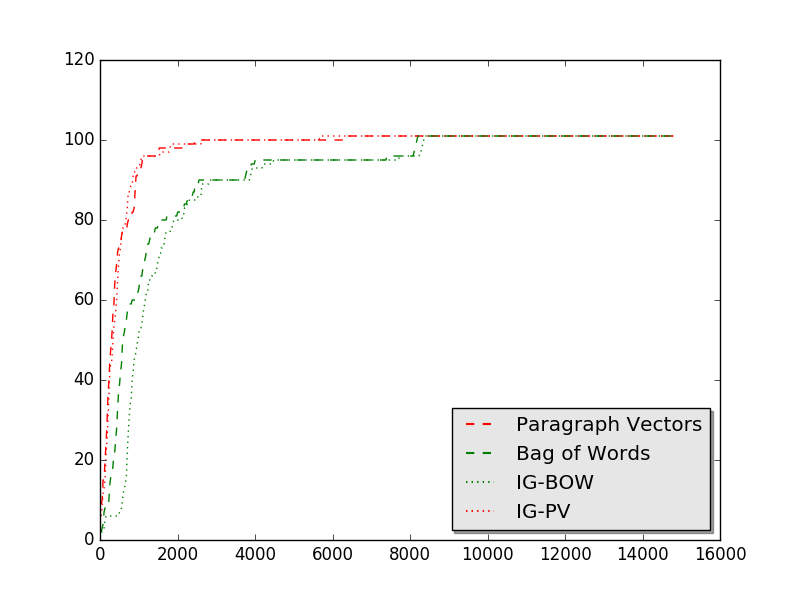}}
\subfloat[LHVS]{\includegraphics[width = 6cm]{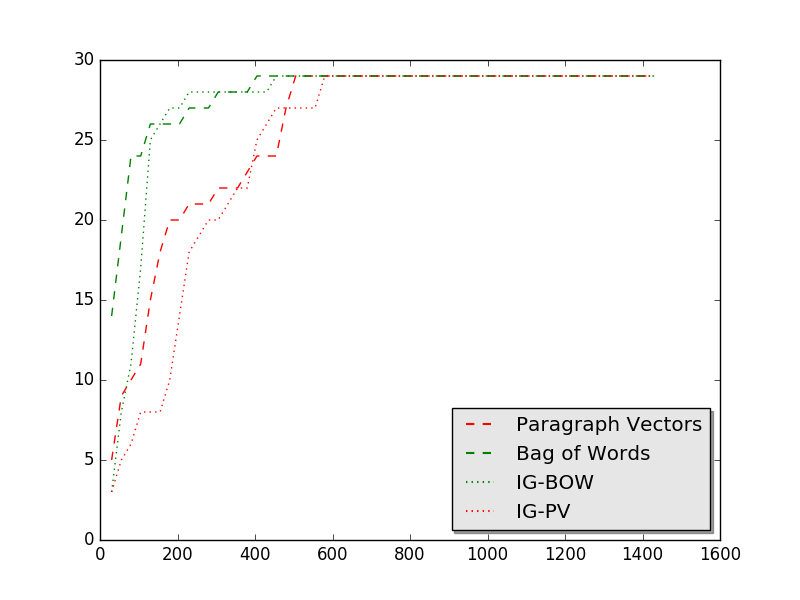}}
\caption{X-axis represents the number of documents that have been manually annotated and Y-axis represents the number of relevant documents that have been discovered. We can observe that our proposed active learning algorithm, that is IG-PV and IG-BOW explore during the initial phases of screening and then their performance improves as the process continues. We show the results of proposed active learning algorithm compared to the naive active learning algorithm in terms of WSS@95 in Figure \ref{fig:bar_chart}.}
\label{fig:iglinegraph}
\end{figure*}

\begin{figure*}
    \centering
    \includegraphics[scale=0.9]{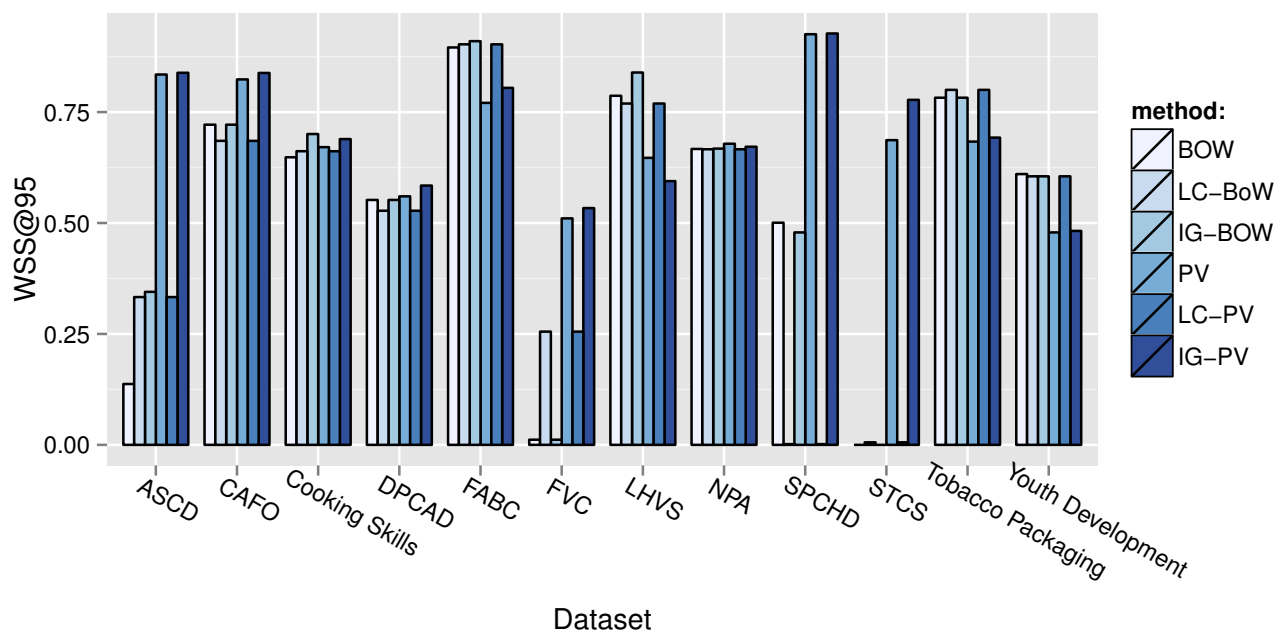}
    \caption{The figure plots the performance of proposed active learning algorithm and the naive active learning algorithm in terms of WSS@95. We use the two best performing feature extraction models (i.e. BoW and PV) for comparing the proposed active learning algorithm with the naive algorithm and baseline algorithm (LC-Bow/LC-PV). We can observe that in 9 out of the 12 datasets the proposed approach (IG-PV/IG-BoW) is the best performing in terms of WSS@95. The winners have a statistically significant lead over all the losers using a t-test at $p<0.05$.}
    \label{fig:bar_chart}
\end{figure*}

When we use our active learning algorithm with BoW model, then we refer to it as IG-BOW (Information Gain-BOW), whereas when we use the paragraph vector model, we refer to it as IG-PV (Information Gain-PV).  We can observe the performance of proposed active learning algorithm compared to the naive active learning algorithm as the screening  progresses in Figure \ref{fig:iglinegraph}. We can notice that our proposed algorithm explores in the initial phases of the screening process, but as the process continues the performance improves. It outperforms naive active learning algorithm by the end of the screening process. We plot the WSS@95 (i.e. WSS at 95\% yield) in Figure \ref{fig:bar_chart} for our proposed active learning algorithm. In addition to the naive active learning algorithm, we compare our method with an additional baseline algorithm that selects the samples about which the classifier is least confident during the initial 10\% (set using cross validation on independent dataset) screening i.e. it randomly selects k studies that have $0.4 \leq p(y=1|d) \leq 0.6$.  We refer to this baseline as LC, and consequently, LC-BoW uses bag-of-words as features and LC-PV uses paragraph vectors as features. We can observe that one of the two, i.e. IG-BoW or IG-PV, performs better compared to naive active learning algorithm using just BoW or PV. It validates our hypothesis that using novelty to explore during the initial phases of active learning can lead to better results overall, especially in terms of WSS@95. We only present BoW and PV because (as we mentioned in the previous paragraph) we observed that these two feature extraction methods perform well across most studies. We should mention at this point that we also used our active learning algorithm with the feature extraction model presented in \cite{hashimoto2016topic}, but it did not perform comparable to PV or BoW. Therefore, we did not present the results with PV-TM to keep the analysis simple and sequential.

It has been shown in our results that both PV and BoW work well for specific systematic reviews. It is not feasible to estimate with certainty in advance which of the two would work better for a review. Meanwhile, we obtained the recall for both feature extraction methods after screening through initial 10\% of the data. We can see that comparison in Table \ref{tab:recall@10}. We report the recall for the initial 10\% of the data and denote in bold the approach that scores higher in terms of WSS@95 by the end of the process. We observe that the approach that has higher recall on the initial 10\% of the data works well in terms of WSS@95 by the end of the screening process. As a result, we infer that both approaches - IG-BOW or IG-PV - should be used in the  beginning and then the one that performs better should be continued.

We should mention that we experimented with different ensembles of feature extraction models, but since the results were not impressive compared to the individual models, we omit the results. 

\begin{tiny}
\begin{table}[ht!]
\label{ref:table}
    \centering
    \begin{tabular}{|c|c|c|}
    \hline\hline
        Dataset & IG-BOW & IG-PV\\ \hline\hline
        LHVS & \textbf{0.896} & 0.275\\ \hline
        ASCD & 0.537 & \textbf{0.671}\\ \hline
        FABC & \textbf{0.957} & 0.899\\ \hline
        NPA &  0.458 & \textbf{ 0.471}\\ \hline
        CAFO &  0.345 & \textbf{0.821}\\ \hline
        DPCAD & 0.403 & \textbf{0.634}\\ \hline
        STCS &  0.322 & \textbf{ 0.613} \\ \hline
        FVC &  0.275 & \textbf{0.345}\\ \hline
        SPCHD & 0.693 & \textbf{0.950}\\ \hline
        Cooking Skills & \textbf{0.497} & 0.738\\ \hline
        Youth Development & \textbf{ 0.435} & 0.426\\ \hline
        Tobacco Packaging & \textbf{ 0.757} & 0.686\\ \hline
    \end{tabular}
    \caption{Recall upon screening through 10\% of the studies in the review. We use the bold notation to mark the overall winner in terms of WSS@95 at the end of the screening process.}
    \label{tab:recall@10}
\end{table}
\end{tiny}

\section{Discussion}
We should mention at this point that we also experimented with learning paragraph vectors using additional external data in the case of public health and animal studies. We expected that additional documents might lead to improved paragraph vectors. These external data consisted of studies related to the review in question, but most of these were not directly relevant to the review. To our surprise, we did not see any improvement in the results upon using such external data for learning paragraph vectors. On the contrary, the performance decreased as we used more external data during the learning of paragraph vectors. We should also mention that we conducted our experiments on more than 30 reviews, and obtained results similar to those presented in this paper. 

We observed that active learning can be an effective strategy to semi-automate manual annotations and reduce the workload. But, the current active learning approaches do not accurately estimate the proportion of relevant studies that have been annotated. In many cases, it is necessary to extract 95\% or more relevant studies in order to avoid bias in the systematic review. If there was an effective way to  estimate the recall of the active learner precisely then the screening process could move towards complete automation. In the future, we will work towards an accurate recall estimation while using an efficient active learning strategy.

\section{Conclusion}
We evaluated different feature extraction models over a comprehensive dataset of reviews from varied domains. We observed that both BoW and PV can outperform other approaches over certain reviews and domains. 

We recognized that a naive active learning algorithm suffers from bias. It tries to select documents that are similar to documents in the training data. Initially, the training data are small and this could lead to reduced performance. We propose a novelty based active learning algorithm that works on exploring different topics during the initial phases of the active learning process and then proceeds based on relevance in the later phases. It leads to more exploration of different topics in the beginning and better performance in terms of WSS@95 by the end. We evaluate our approach against naive active learning algorithm, and observe that the proposed algorithm works equal to, or better than, the naive algorithm in all instances.

We also develop insights regarding the choice of feature extraction methods for different reviews. We observe based on our experiments on a large number of reviews, that both the feature extraction methods should be used during the screening of initial 10\% studies. Afterwards, the better performing approach should be continued. It leads to some extra manual annotations in the beginning, but the overall gain in terms of WSS@95 compensates for that disadvantage.

\section{Acknowledgement}
We acknowledge with thanks Annette O'Connor (Iowa State University), the Cochrane Heart Group and the EPPI-Centre (University College London) review team who supplied data for conducting this study. This work was supported by a grant awarded by the UK Medical Research Council: Identifying relevant studies for systematic reviews and health technology assessments using text mining [Grant No. MR/J005037/1]. James Thomas was (in part) supported by the National Institute for Health Research (NIHR) Collaboration for Leadership in Applied Health Research and Care (CLAHRC) North Thames at Bart's Health  NHS Trust. The views expressed are those of the author(s) and not necessarily those of the NHS, the NIHR or the Department of Health.

\bibliographystyle{abbrv}
\bibliography{biblio}
\end{document}